\documentclass[preprint,12pt]{elsarticle}

\usepackage{amssymb}
\usepackage{amsfonts}
\usepackage{amsmath}
\usepackage{graphicx}
\usepackage{bm}
\usepackage{braket}

\usepackage{url}

\begin{document}

\title{Bessel Vortices in Spin-Orbit-Coupled Binary Bose-Einstein Condensates with Zeeman Splitting}
\author[inst1]{Huan-Bo Luo}
\author[inst2]{B. A. Malomed}
\author[inst3]{Wu-Ming Liu}

\affiliation[inst1]{organization={Institute of Theoretical Physics and Department of Physics, State Key
Laboratory of Quantum Optics and Quantum Optics Devices, Collaborative
Innovation Center of Extreme Optics},%Department and Organization
            addressline={Shanxi University}, 
            city={Taiyuan 030006},
            country={China}
}

\affiliation[inst2]{organization={Department of Physical Electronics, School of Electrical Engineering,
Faculty of Engineering},%Department and Organization
            addressline={Tel Aviv University}, 
            city={Tel Aviv 69978},
            country={Israel}
}

\affiliation[inst3]{organization={Beijing National Laboratory for Condensed Matter Physics, Institute of Physics},%Department and Organization
            addressline={Chinese Academy of Sciences}, 
            city={Beijing 100190},
            country={China}
}
\author[inst1]{Lu Li\corref{cor1}}
 \ead{llz@sxu.edu.cn}
 \cortext[cor1]{Corresponding author.}

\begin{abstract}
We present an analysis of stationary solutions for two-dimensional (2D) Bose-Einstein
condensates (BECs) with the Rashba spin-orbit (SO) coupling and Zeeman
splitting. By introducing the generalized momentum operator, the linear
version of the system can be solved exactly. The solutions are semi-vortices
of the Bessel-vortex (BV) and modified Bessel-vortex (MBV) types, in the presence of
the weak and strong Zeeman splitting, respectively. The ground states (GSs) of the 
full nonlinear system are constructed with the help of a specially designed neural 
network (NN). The GS of the mixed-mode type appears as cross-attraction interaction increases. 
The spin texture of the GS is produced in detail. It exhibits the 
N\'{e}el skyrmion structure for the semi-vortex GS of the BV type, and the 
respective skyrmion number is found in an analytical form. On the other hand, 
GSs of the MBV and mixed-mode types do not form skyrmions.
\end{abstract}

\begin{keyword}
    Bessel vortices \sep Spin-orbit  coupling \sep Zeeman splitting \sep Neural network
\end{keyword}

\maketitle

%\pacs{03.75.Mn, 05.30.Jp, 03.75.Lm}

\section{Introduction}

Atomic Bose-Einstein condensates (BECs) are perfectly controllable quantum
settings, which make it possible to emulate various effects which are well
known in condensed-matter systems~\cite{RepProgPhys.75.082401,Lewenstein}. A
well-known example is the spin-orbit (SO) coupling in semiconductors,
originating from the interaction of the electron spin with the weak magnetic
field induced as the Lorentz transform of the electrostatic field of the
crystalline lattice in the reference frame moving along with the electron.
The solid-state SO coupling plays a fundamental role in the realization of
spin Hall effects~\cite{RevModPhys.82.1959}, topological insulators~\cite%
{RevModPhys.82.3045}, spintronic devices~\cite{RevModPhys.76.323}, etc.
There are two standard forms of the Hamiltonian of the SO coupling, viz.,
ones of the Dresselhaus~\cite{Dresselhaus} and Rashba~\cite{Rashba,Sherman}
types.

The last decade has witnessed the experimental realization of the emulation
of the SO coupling in effectively one-~\cite{nature09887,Juzeliunas} and
two-dimensional~(2D)~\cite{Science.354.83} BEC, see also reviews of the
experimental and theoretical results in Refs.~\cite%
{Spielman,Galitski,Ohberg,Zhai}. Simultaneously, using Gross-Pitaevskii
equations modeling the system~\cite{BEC-SOC GP eqns}, many remarkable
effects have been predicted in SO-coupled BECs with intrinsic nonlinearity.
These include vortices~\cite%
{Kawakami,Drummond,PhysRevLett.109.015301,Sakaguchi}, skyrmions~\cite%
{PhysRevLett.109.015301} and many species of 1D~\cite%
{PhysRevLett.110.264101,1D sol 2,1D sol 3,1D sol 4}, 2D~\cite%
{PhysRevE.89.032920,PhysRevE.94.032202,Cardoso,Lobanov,2D SOC gap sol
Raymond,SOC 2D gap sol Hidetsugu,low-dim SOC}, and 3D~\cite{Han Pu 3D}
solitons, see also review~\cite{SOC-sol-review}. Further, the SO coupling
makes it possible to predict novel states of matter, such as chiral
supersolids~\cite{PhysRevLett.121.030404} and polariton topological
insulators~\cite{PhysRevLett.122.083902}.

In this paper, we investigate stationary solutions for binary BEC with the
SO coupling and Zeeman splitting between its two components. Starting from
exact solutions of the linear version of the SO-coupled Gross-Pitaevskii
equations, represented by Bessel vortices (BV) and modified Bessel vortices
(MBV) in the presence of weak and strong Zeeman effects, respectively, we
develop a neural network (NN) for simulating the system with the intrinsic
attractive nonlinearity. 
The NN based numerical method \cite{IEEE95,GAMM} is inherently parallel 
and even large-scale system can easily
get accelerated with the mordern Artificial Intelligence based
computer. In this study, both the ground and
excited states can be obtained by NN. 
In particular, the NN solutions produce 2D (\textquotedblleft baby"~\cite%
{baby1,baby,Shnir}) skyrmions in the SO-coupled BEC (note that a baby
skyrmion in a three-component condensate was created experimentally \cite%
{Choi}). The corresponding values of the skyrmion number can be calculated
analytically. Stability of the so predicted states is checked by dint of
numerical simulations.

We consider a binary SO-coupled $^7$Li BEC with attractive contact interaction and
the Zeeman splitting between its components in the 2D space. SO coupling is created by
laser beams which couple different states of $^7$Li atoms~\cite{Lett.108.035302}. The spinor wave
function, $\Psi =(\Psi _{1},\Psi _{2})^{T}$, of this system is governed by
the system of Gross-Pitaevskii equations. The scaled form~of these equations
is \cite{PhysRevE.94.032202,PhysRevLett.122.123201}
\begin{equation}
    \begin{split}
i\partial _{t}\Psi _{1} &=-\frac{1}{2}\nabla _{\bot }^{2}\Psi _{1}+\beta %
\left[ (\partial _{x}-i\partial _{y})\Psi _{2}+k_{z}\Psi _{1}\right]  
-\left( g|\Psi _{1}|^{2}+\gamma |\Psi _{2}|^{2}\right) \Psi _{1},  \\
i\partial _{t}\Psi _{2} &=-\frac{1}{2}\nabla _{\bot }^{2}\Psi _{2}+\beta %
\left[ -(\partial _{x}+i\partial _{y})\Psi _{1}-k_{z}\Psi _{2}\right]-
\left( \gamma |\Psi _{1}|^{2}+g|\Psi _{2}|^{2}\right) \Psi _{2},\label{main}
    \end{split}
\end{equation}%
and the characteristic length, energy and time are defined by $l=2\mu \text{m},
\epsilon=\hbar^2/ml^2=2.4\times 10^{-31} \text{J}$
and $\tau=\hbar/\epsilon=0.44\text{ms}$~\cite{PhysRevLett.122.123201}, where  
$m=1.16\times10^{-26}$kg is the mass of $^7$Li.
The operator $\nabla _{\bot }^{2}$ acts in the 2D plane $\left( x,y\right)
$. The coefficient $\beta=1 $ is the SO coupling strength which can be varied in a broad range
depending on laser configurations~\cite{Lett.108.035302}.  $g=1$ and $\gamma=a_\gamma/a_g $ are
coefficients of the self- and cross-attraction interaction, respectively.
$a_g$ and $a_\gamma$ are the $s$-wave scattering lengths between atoms in same states and different states.
In order to maintain the generality of the system, both $a_g$ and $a_\gamma$ can be varied in this system.
The total number of atoms $U=(N/a_g)\sqrt{\hbar/8\pi m \omega_z}$,
where $N=\int \Psi^\dagger\Psi dxdy$ is the norm integral and  $\omega_z=800$Hz~\cite{Lett.87.130402} is the frequency of tight confinement
along the $z$ axis.
The Zeeman-splitting strength is $\Omega=\mu_Bg_FB/\epsilon =\beta k_{z}$, where $k_{z}$
may be considered as the $z$ component of the momentum, $g_F$ is the Land\'{e} g factor for $^7$Li, 
$\mu_B$ is the Bohr magneton and $B$ is a uniform magnetic field in $z$ axis. We can find that only $k_{z}$ and $\gamma$ are free
parameters of the system.

Stationary solutions of Eq.~\eqref{main} with chemical potential $\mu $ are
sought for in the usual form,
\begin{equation}
\Psi =\{\psi _{1}(x,y),\psi _{2}(x,y)\}^{T}\exp (-i\mu t).  \label{Psipsi}
\end{equation}%
with stationary functions $\psi _{1,2}(x,y)$ satisfying equations%
\begin{equation}
    \begin{split}  
\mu \psi _{1} &=-\frac{1}{2}\nabla _{\bot }^{2}\psi _{1}+
(\partial _{x}-i\partial _{y})\psi _{2}+k_{z}\psi _{1} 
-\left( |\psi _{1}|^{2}+\gamma |\psi _{2}|^{2}\right) \psi _{1},  \\
\mu \psi _{2} &=-\frac{1}{2}\nabla _{\bot }^{2}\psi _{2}
-(\partial _{x}+i\partial _{y})\psi _{1}-k_{z}\psi _{2} 
-\left( \gamma |\psi _{1}|^{2}+|\psi _{2}|^{2}\right) \psi _{2}, 
    \end{split}\label{main2}
\end{equation}%
In the presence of the Zeeman splitting, Eq. (\ref{main2}) yields two
uniform states, \textit{viz}.,
\begin{equation}
\psi _{1}=C,\psi _{2}=0,\mu =k_{z}-C^{2};  \label{const1}
\end{equation}
\begin{equation}
\psi _{1}=0,\psi _{2}=C,\mu =-k_{z}-C^{2},  \label{const2}
\end{equation}
where $C$ is a real constant. In this paper, we only focus on the branch
with lower chemical potential $\mu $, as it is plausible that the higher
branch is unstable. As concerns the uniform states, the lower branch
corresponds to Eq. (\ref{const2}), in the case of $k_{z}>0$.

A family of vortex solutions is defined by the ansatz~\cite{PhysRevE.89.032920,PhysRevE.94.032202} which is compatible
with the substitution of expression (\ref{Psipsi}) in Eq. (\ref{main}):
\begin{equation}
\psi _{1}(x,y)=R_{1}(r)e^{-i(m+1)\theta },\psi
_{2}(x,y)=R_{2}(r)e^{-im\theta },  \label{vortex}
\end{equation}%
where $\left( r,\theta \right) $ are the polar coordinates, $m$ is an
integer winding number, and $R_{1,2}(r)$ are two radial wave functions. The
simplest (fundamental) version of ansatz (\ref{vortex}), corresponding to $%
m=0$, or its mirror image with $m=-1$, represents the ground state (GS), in
the form of the \textit{semi-vortex}, as defined in Ref. \cite{Sakaguchi}
(see also review \cite{PhysicaD}), and the ones with \textit{excitation
number} $m\geq 1$ or $m\leq -2$ are defined, in the same works, as \textit{%
excited states} of the semi-vortex unlike the GS, the excited states were
found to be completely unstable (in the absence of the Zeeman splitting)
\cite{Sakaguchi}, therefore only the GS solutions are considered here. Note
that Eq.~\eqref{main} is compatible with substitution $k_{z}\rightarrow
-k_{z}$, $m\rightarrow -m-1$, $R_{1}\rightarrow -R_{2}$, $R_{2}\rightarrow
R_{1}$, therefore it is sufficient to consider only positive values of $%
k_{z} $.

\section{Exact vortex states of the linearized system}

First, we note that the~stationary linear version of Eq.~\eqref{main}, i.e.,
$\hat{H}\psi =\mu \psi $ with Hamiltonian
\begin{equation}
\hat{H}=-\nabla_{\bot } ^{2}/2-(i\sigma _{x}\partial _{y}-i\sigma _{y}\partial
_{x}-k_{z}\sigma _{z}),  \label{H}
\end{equation}%
where $\sigma _{x,y,z}$ are the Pauli matrices, admits an exact solution.
Indeed, in terms of the generalized momentum operator,
\begin{equation}
\hat{P}=i\sigma _{x}\partial _{y}-i\sigma _{y}\partial _{x}-k_{z}\sigma _{z},
\label{P}
\end{equation}%
Hamiltonian (\ref{H}) can be written as $H=\hat{P}^{2}/2-\hat{P}-k_{z}^{2}/2$%
. Then, solving the eigenvalue equation $\hat{P}\psi =k\psi $ with real $k$
yields the exact solution of the linearized system (\ref{main}), written in
terms of the Bessel functions:
\begin{equation}
R_{1}(r)=Ck_{r}J_{m+1}(k_{r}r),R_{2}(r)=C(k+k_{z})J_{m}(k_{r}r),
\label{psi_1}
\end{equation}%
where $C$ is an arbitrary real constant [similar to that in Eqs. (\ref%
{const1}) and (\ref{const2})], $k_{r}$ is the radial momentum, and the total
momentum $k$ is defined by
\begin{equation}
k^{2}=k_{r}^{2}+k_{z}^{2}.  \label{kkrkz}
\end{equation}%
The solution is built as a pair of the Bessel Vortices (BV) with winding
numbers $-(m+1)$ and $-m$ in the $\psi _{1}$ and $\psi _{2}$ components,
respectively. In the absence of the Zeeman effect, i.e., $k_{z}=0$, this
solution reduces to one that was recently reported in Ref.~\cite{Viskol}.
Naturally, the norm integral for this linear state in the free space
diverges as%
\begin{equation}
N\equiv N_{1}+N_{2}=\lim_{R\rightarrow \infty }\left\{ 2\pi \int_{0}^{R}
\left[ R_{1}^{2}(r)+R_{2}^{2}(r)\right] rdr\right\}   
\simeq 4C^{2}\frac{k\left( k+k_{z}\right) }{k_{r}}R.  \label{norm}
\end{equation}%
while the ratio of the norms of the two components is finite:%
\begin{equation}
N_{1}/N_{2}=k_{r}^{2}/\left( k+k_{z}\right) ^{2}.  \label{N/N}
\end{equation}

Because operator $\hat{P}$, defined by Eq. (\ref{P}), commutes with $\hat{H}$%
, the state $\psi $ given by Eq.~\eqref{psi_1} is also the eigenstate of $%
\hat{H}$, with the respective chemical potential
\begin{equation}
\mu (k)=k^{2}/2-k-k_{z}^{2}/2,  \label{mu}
\end{equation}%
where the three terms represent, severally, the kinetic energy, SO coupling,
and Zeeman energy shift. In the case of $k_{z}<1$, the chemical
potential attains its minimum,
\begin{equation}
\mu _{\min }(k_{z}<1)=-1/2-k_{z}^{2}/2,  \label{min mu}
\end{equation}%
at $k=1$, hence Eq. (\ref{kkrkz}) gives, in this case,
\begin{equation}
k_{r}=\sqrt{1-k_{z}^{2}}.  \label{kr}
\end{equation}%
In the limit of $k_{z}=1$ and $k_{r}=0$, the BV solution (\ref{psi_1}) amounts
to the uniform state, with $R_{1}=0$, $R_{2}=2C$ for $m=0,$ and to zero
solution for $m\geq 1$.

At $k_{z}>1$, Eq. (\ref{kr}) yields an imaginary radial momentum $k_{r}$,
and the BV solution (\ref{psi_1}) becomes the \textit{modified Bessel Vortex}
(MBV)
\begin{equation}
R_{1}=-C|k_{r}|K_{m+1}(|k_{r}|r),R_{2}=C(k+k_{z})K_{m}(|k_{r}|r),
\label{MBV}
\end{equation}%
where $K_{m}$ is the standard modified Bessel function of the second kind
(alias the modified Hankel function), which exponentially decays at $%
r\rightarrow \infty $,
\begin{equation}
K_{m}(|k_{r}|r)\underset{r\rightarrow \infty }{\approx }\sqrt{\frac{\pi }{%
2|k_{r}|r}}\exp \left( -|k_{r}|r\right)  \label{exp}
\end{equation}%
but is singular at $r\rightarrow 0$,
\begin{equation}
K_{m}(|k_{r}|r)\underset{r\rightarrow 0}{\approx }\frac{1}{2}(|m|-1)!\left(
\frac{2}{|k_{r}|r}\right) ^{|m|}  \label{m>=1}
\end{equation}%
for $m\geq 1$, and
\begin{equation}
K_{0}(|k_{r}|r)\underset{r\rightarrow 0}{\approx }\ln \left( \frac{2}{%
|k_{r}|r}\right) .  \label{m=0}
\end{equation}%
Accordingly, the norm (\ref{norm}) of the MBV solutions of the linearized
system diverges at $r\rightarrow 0$ as $r^{-2(m+1)}$ for $m\geq 1$, and as $%
\ln (1/r)$ for $m=0$, while the ratio of the norms of the two components
vanishes in the same limit.

As shown below, taking into account the self-attractive nonlinearity in Eq. (%
\ref{main2}) makes it possible to replace the GSs of the BV and MBV states
by similar ones, but with a finite norm. Note that the nonlinearity does 
not affect the exponentially decaying asymptotic expression (\ref{exp}).

Finally, in the case of $k_{z}>1$, the minimum value of the chemical
potential is achieved at $k_{r}=0$, hence Eqs. (\ref{kkrkz}) and (\ref{mu})
yield%
\begin{equation}
\mu_{\min }\left( k_{z}>1\right) =-k_{z}.  \label{min mu 2}
\end{equation}

The results show that, with the increase of the Zeeman-splitting strength $%
k_{z}$, the minimum chemical potential, given by Eq. (\ref{min mu}),
decreases, as per Eqs. (\ref{min mu}) and (\ref{min mu 2}), and the
corresponding wave function carries over from the BV state ($0\leq k_{z}<1$)
to the MBV one ($k_{z}>1$). This is, in fact, an example of the quantum
phase transition, as it occurs in terms of the mean-field wave function of
the quantum gas, cf. Refs. \cite{Ueda,Sengstock,Dutta}. A similar phase
transition between BV and MBV localized states, with finite norms, is
reported below. Note that all the vortex states are degenerate with respect
to the excitation number $m$, as $\mu $, given by Eq.~\eqref{mu}, does not
depend on $m$.

\section{Constructing nonlinear vortex states by means of the neural network}

Based on the exact vortex states found for the linear version, we aim to
construct their nonlinear counterparts, as solutions of Eq. (\ref{main}),
with the help of a neural network (NN). First, we consider the case of $%
\gamma =0$ [no nonlinear interaction between the components of the wave
function in Eq. (\ref{main}). Substituting the general ansatz~\eqref{vortex}
in Eq.~\eqref{main} with $\gamma =0$ yields
\begin{equation}
G_{1,2}(r,R_{1},R_{1}^{\prime },R_{1}^{\prime \prime },R_{2},R_{2}^{\prime
},R_{2}^{\prime \prime })=0,  \label{radial}
\end{equation}%
where%
\begin{equation}
\begin{split}
G_{1}=& \left( -\mu +k_{z}\right) R_{1}-\frac{1}{2}\left[ R_{1}^{\prime
\prime }+\frac{R_{1}^{\prime }}{r}-\frac{(m+1)^{2}R_{1}}{r^{2}}\right] 
+R_{2}^{\prime }-\frac{mR_{2}}{r}-R_{1}^{3}, \\
G_{2}=& \left( -\mu -k_{z}\right) R_{2}-\frac{1}{2}\left[ R_{2}^{\prime
\prime }+\frac{R_{2}^{\prime }}{r}-\frac{m^{2}R_{2}}{r^{2}}\right] 
 -R_{1}^{\prime }-\frac{(m+1)R_{1}}{r}-R_{2}^{3}.
\end{split}
\label{12}
\end{equation}%
Next, introducing the error index,
\begin{equation}
L\equiv 2\pi \int_{0}^{\infty }r\left( G_{1}^{2}+G_{2}^{2}\right) dr,
\label{L}
\end{equation}%
the task is to drive its value to zero as close as possible for given
chemical potential $\mu $.

\begin{figure}[tbp]
\centering
%Requires \usepackage{graphicx}
\includegraphics[width=5 in]{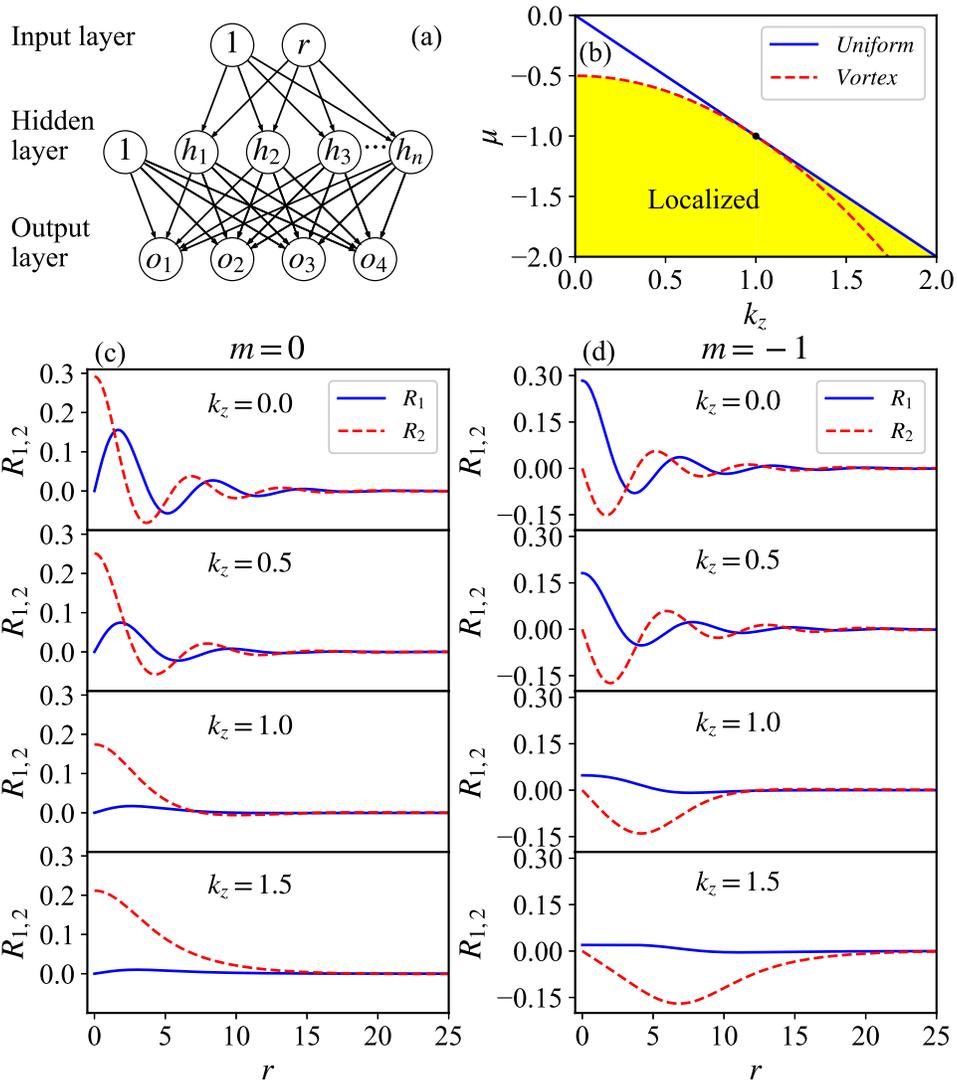}
\caption{(Color online) (a) The schematic of the NN with $1+1$ input nodes, $%
1+n$ hidden nodes, and $4$ output nodes. (b) The linear dispersion spectrum
of the system. Localized states exist in the yellow area. Profiles of the
radial wave functions $R_{1,2}$ are displayed in panels (c) for $m=0$ and
(d) for $m=-1$, at $k_{z}=0.0$, $0.5$, $1$ and $1.5$. Here $\Delta =0.01$
[see Eq. (\protect\ref{Delta})].}
\label{figure1}
\end{figure}

In the nonlinear system, localized solutions may exist in the \textit{bandgap%
} of the linear spectrum of Eq. (\ref{main2}), i.e., at values of $\mu $ at
which the linear solutions (\ref{psi_1}) and (\ref{MBV}) do not exist, cf.
Ref.~\cite{PhysRevE.94.032202}. As it follows from Eqs. (\ref{min mu}) and (%
\ref{min mu 2}), the bandgap is%
\begin{equation}
\mu <\left\{
\begin{array}{c}
-1/2-k_{z}^{2}/2,~\mathrm{at}~k_{z}<1, \\
-k_{z},~\mathrm{at}~k_{z}>1.%
\end{array}%
\right.  \label{bandgap}
\end{equation}
It is shown by the yellow area in the plane of $\left( k_{z},\mu \right) $,
in Fig.~\ref{figure1}(b). Thus, the chemical potential of localized states
can be denoted as
\begin{equation}
\mu =\mu _{\min }-\Delta ,~\mathrm{with}~~\Delta >0.  \label{Delta}
\end{equation}

As the self-focusing nonlinearity can chop off the slowly decaying tails of
the Bessel wave function, which makes its norm integral norm diverging [see
Eq. (\ref{norm})], we adopt the following ansatz, which agrees with the
general structure of the BV solutions \eqref{psi_1} and includes the
truncation factor $\mathrm{sech}(ar)$:
\begin{equation}
R_{1}=o_{1}\mathrm{sech}(ar)J_{m+1}(o_{3}r),R_{2}=o_{2}%
\mathrm{sech}(ar)J_{m}(o_{4}r),  \label{bessel}
\end{equation}%
where $a=\sqrt{2\Delta }$ is an empirical choice, suggested by numerical
computations [recall $\Delta $ is defined by Eq. (\ref{Delta})]. Both BV and
MBV states can be produced by the input taken as per ansatz (\ref{bessel}).

A schematic of the NN used in this study is shown in Fig.~\ref{figure1}(a).
It includes three layers, \textit{viz}., an input one $X=[1,r]^{T}$, hidden
layer $H_{l}=[1,h_{1},h_{2},h_{3},\cdots ,h_{n}]^{T}$, and the output one, $%
O=[o_{1},o_{2},o_{3},o_{4}]^{T}$. The appended ``1" in $X$ and $H_{l}$ is
used to account for the bias in the input and hidden layer. 
The bias has the effect of shifting the activation function by a constant, 
which can improve the accuracy of output generated by the NN. The
hidden and output layers can be expressed as
\begin{equation}
H_{l}=[1,f(WX)^{T}]^{T},O=Af(VH_{l}),  \label{Hl}
\end{equation}%
where $W$ and $V$ are $2\times n$ and $(n+1)\times 4$ parameter matrices.
Here, we choose $n=12$ and $A=2$, initial values of $W$ and $V$ are given by
random numbers uniformly distributed between $-1$ and $+1$, and $f(x)=1/(1+e^{-x})$ is a
component-wise sigmoid activation function. A component-wise function $y=f(x)$ 
can be defined as that if $x$ is matrix, then $y_{ij}=f(x_{ij})$. Thus, for given $W$ and
$V$, we can obtain an error index $L(W,V)$ and its gradient with respect to $%
W$ and $V$. Based on the gradient, the parameter matrices $W$ and $V$ can be
updated by means of the adaptive-moment estimation (Adam) method. Note that
this method cannot avoid producing the trivial solution, $R_{1}=R_{2}=0$, of
Eq.~\eqref{radial}. In that case, the total norm quickly decays toward zero,
and we then need to generate a new set of $W$ and $V$ for the next
iteration. The soliton solution can be found if the total norm becomes a
constant instead of decaying to zero.

\begin{figure}[tbp]
    \centering
    %Requires \usepackage{graphicx}
    \includegraphics[width=5in]{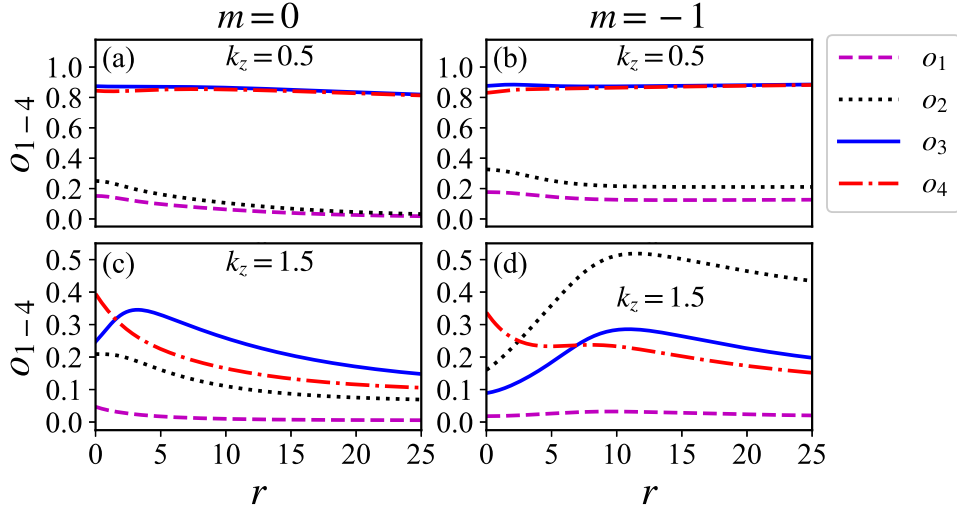}
    \caption{(Color online) Profiles of the $o_{1-4}$ functions generated by NN are displayed in 
    panels (a,c) for $m=0$ and
    (b,d) for $m=-1$, at $k_{z}=0.5$ and $1.5$. Here $\Delta =0.01$}
    \label{figure2}
\end{figure}

As mentioned above, we focus on the consideration of the GSs, which are
represented by ansatz (\ref{psi_1}) with $m=0$ and $-1$. The BV ($0\leq
k_{z}<1$) and MBV ($k_{z}>1$) states with $m=0,-1$ were calculated at $%
\Delta =0.01$, as shown in Figs.~\ref{figure1}(c-d). It takes about $3\times
10^{4}$ iterations to reduce values of the error index (\ref{L}) to $%
L<10^{-6}$. Naturally, the radial wave functions of the BV states feature
the oscillatory decaying tail, while the oscillations are absent in the MBV
states. Because the presence or absence of the oscillations in the tails of
the localized states is the indicator of the quantum phase transition
between ones of the BV and MBV types, $k_{z}=1$ remains the phase-transition
point, exactly the same as in the linear system considered above. It is
relevant to mention that, although the system (\ref{main}) dealt with here
is similar to one addressed in Ref. \cite{PhysRevE.94.032202}, the BV $%
\rightarrow $ MBV phase transition at $k_{z}=1$ was not considered in that
work. Profiles of the $o_{1-4}$ functions generated by NN, as intermediate results, 
are shown in Fig.~\ref{figure2}.

\begin{figure}[tbp]
    \centering
    %Requires \usepackage{graphicx}
    \includegraphics[width=5in]{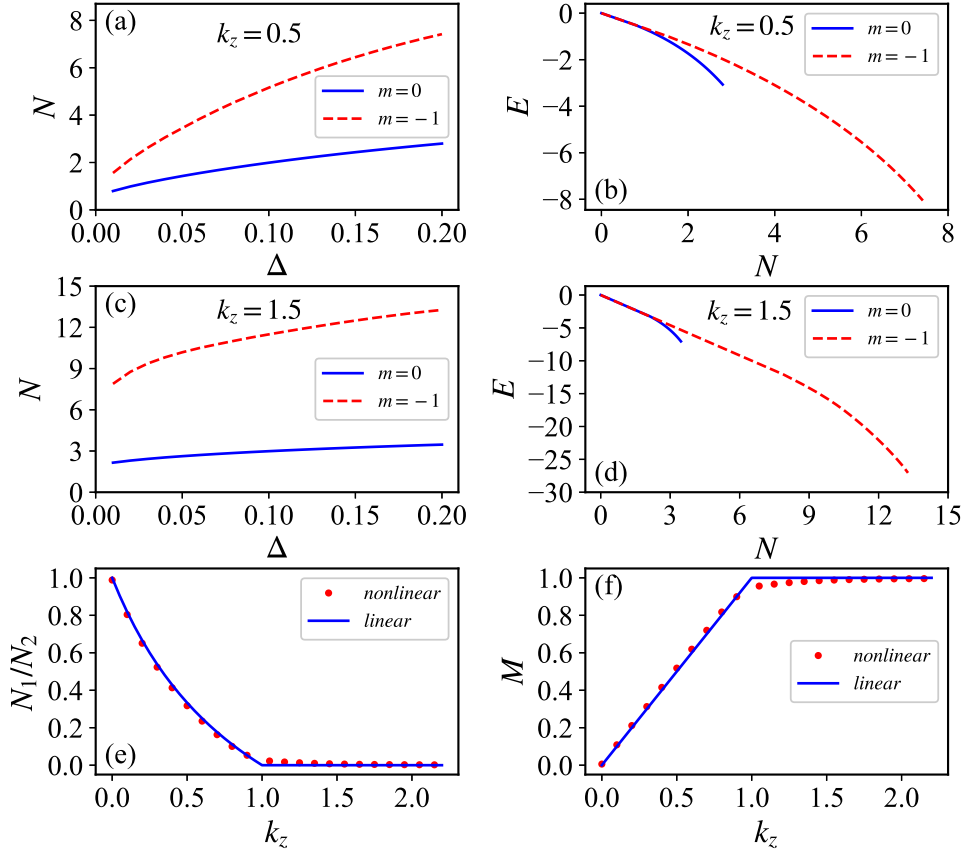}
    \caption{(Color online) The total norm $N$ as a function of $\Delta $ [see
    Eq. (\protect\ref{Delta})] for the states with $m=0$ and $-1$ and (a) $%
    k_{z}=0.5$ and (c) $k_{z}=1.5$. The corresponding total energy (\protect\ref%
    {E}) for (b) $k_{z}=0.5$ and (d) $k_{z}=1.5$. (e) Ratio $N_{1}/N_{2}$ and
    (f) the magnetization curve for the system's GS with $m=0$. Here, the data
    points are calculated at $\Delta =0.01$.}
    \label{figure3}
\end{figure}

The total norm $N$ of the so obtained solutions are plotted, vs. $\Delta $,
in Figs.~\ref{figure3}(a,c). The stability of the stationary solutions can
be identified through the computation of eigenvalues for small perturbations
added, in the linear approximation, to the stationary states. It has thus
been checked that all the GS solutions obtained in this work [for $\gamma =0$
in Eq. (\ref{main})] are stable.

The true GS of the system can be identified by calculating the energy
corresponding to Eq. (\ref{main}),%
\begin{eqnarray}
E &=&\iint  \left\{ \frac{1}{2}\left( \left\vert \nabla \psi
_{1}\right\vert ^{2}+\left\vert \nabla \psi _{2}\right\vert ^{2}\right)
+ \psi _{1}^{\ast }\left( \partial _{x}-i\partial _{y}\right) \psi
_{2}-\psi _{2}^{\ast }\left( \partial _{x}+i\partial _{y}\right) \psi
_{1}\right.  \notag \\
&&\left. +k_{z}\left( \left\vert \psi _{1}\right\vert ^{2}-\left\vert \psi
_{2}\right\vert ^{2}\right) -\frac{g}{2}\left( \left\vert \psi
_{1}\right\vert ^{4}+\left\vert \psi _{2}\right\vert ^{4}\right) -\gamma
\left\vert \psi _{1}\right\vert ^{2}\left\vert \psi _{2}\right\vert
^{2}\right\} dxdy,  \label{E}
\end{eqnarray}%
where $\ast $ stands for the complex conjugate. From Figs.~%
\ref{figure3}(b,d) it is seen that energy of the BV and MBV states with $m=0$
is always lower than that of their counterparts with $m=-1$, which means
that the state with $m=0$ represent the GS. Thus, the introduction of the
Zeeman splitting lifts the degeneracy between the states with $m=0$ and $-1$
\cite{Sakaguchi}.

Further, ratio $N_{1}/N_{2}$ of the norms of the $\psi _{1}$ and $\psi _{2}$
components of the GS with $m=0$ is plotted in Fig.~\ref{figure3}(e). It is
close to the same ratio found for the exact solutions of the linearized
equations, given by Eq. (\ref{N/N}) at $\Delta =0.01$. The two-component
Bose gas can be considered as a (pseudo-)spin system. The spin vector, $%
\mathbf{S}=\psi ^{\dagger }\bm{\sigma}\psi /\psi ^{\dagger }\psi $, can be
used to represent the respective pseudo-magnetic ordering, where $\bm{\sigma}%
=(\sigma _{x},\sigma _{y},\sigma _{z})$ is the Pauli matrix vector. The
average $z$-component of the spin $\bar{S}_{z}=\left\langle \psi \right\vert
\sigma _{z}\left\vert \psi \right\rangle /N=(N_{1}/N_{2}-1)/(N_{1}/N_{2}+1)$%
, and the magnetization is defined as $M=|\bar{S}_{z}|$. Note that $M=k_{z}$
for $0\leq k_{z}<1$ in linear regime. The magnetization curve for $m=0$ is
plotted in Fig.~\ref{figure3}(f). Without the external magnetic field
applied to the BEC, i.e., at $k_{z}=0$, the atoms are evenly distributed in
the $\psi _{1}$ and $\psi _{2}$ components, i.e. $N_{1}/N_{2}=1$, hence the
magnetization is negligibly small. With the increase of the external
magnetic field $k_{z}$, the atomic population is transferred from $%
\psi _{1}$ to $\psi _{2}$, which yields a higher magnetization. Eventually,
at the critical value of the effective field, $k_{z}=1$, nearly all the
atoms are transferred to $\psi _{2}$. The magnetization remains
nearly constant as $k_{z}>1$, which means saturation of the
magnetization.

The above consideration was performed for $\gamma =0$ in Eq. (\ref{main}).
In the presence of $\gamma $, i.e., the nonlinear interaction between the
two components of the wave function, one may expect that the GS is
represented not by the semi-vortex of the BV or MBV type, but by the mixed
mode, in which both components contain mixtures of azimuthal harmonics with
winding numbers $0$ and $\pm 1$, cf. Eq. (\ref{vortex}). Here, we consider
the mixed modes, the initial guess for which is built as a superposition of
semi-vortices of the BV or MBV types, with $m=0$ and $m=-1$:%
\begin{equation}
\begin{split}
\psi _{1}& =\mathrm{sech}(ar)\left[ o_{1}e^{-i\theta
}J_{1}(o_{2}r)+o_{3}J_{0}(o_{4}r)\right] , \\
\psi _{2}& =\mathrm{sech}(ar)\left[ o_{5}J_{0}(o_{6}r)+o_{7}e^{i\theta
}J_{-1}(o_{8}r)\right] ,
\end{split}
\label{MM}
\end{equation}%
cf. Eq. (\ref{bessel}). The weight of each semi-vortex in this ansatz may be
defined as
\begin{equation}
c_{0}=d_{0}/(d_{0}+d_{-1}),~c_{-1}=1-c_{0},  \label{cc}
\end{equation}%
where
\begin{equation}
\begin{split}
d_{0}& =\iint \mathrm{sech}^{2}(ar)\left[
o_{1}^{2}J_{1}^{2}(o_{2}r)+o_{5}^{2}J_{0}^{2}(o_{6}r)\right] dxdy, \\
d_{-1}& =\iint \mathrm{sech}^{2}(ar)\left[
o_{3}^{2}J_{0}^{2}(o_{4}r)+o_{7}^{2}J_{-1}^{2}(o_{8}r)\right] dxdy.
\end{split}
\label{c0}
\end{equation}%
are norms of both semivortex constituents. It follows from Eqs. (\ref{cc})
and (\ref{c0}) that weights $c_{0}$ and $c_{-1}$ belong to the interval of $%
0<c_{0},c_{-1}<1$.

\begin{figure}[tbp]
\centering
%Requires \usepackage{graphicx}
\includegraphics[width=5in]{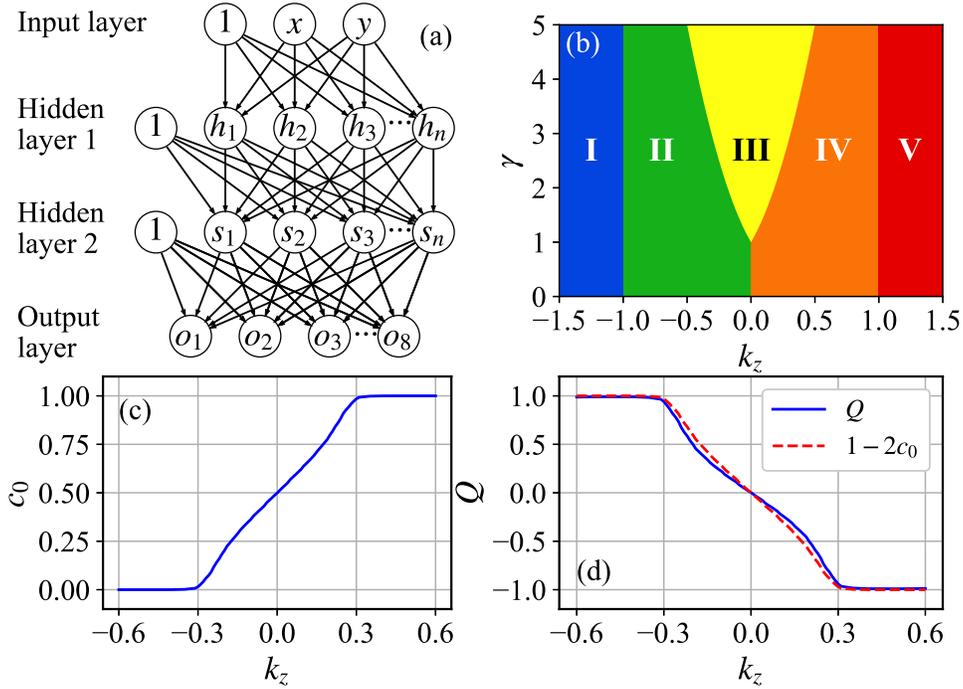}
\caption{(Color online) The schematic of the NN with $2+1$ input nodes, two
layers of $n+1$ hidden nodes, and $8$ output ones. (b) Diagram of the GS of
the MBV semi-vortex type with (I) $m=-1$ and (V) $m=0$; the GS of the BV
semi-vortex type with (II) $m=-1$ and (IV) $m=0$; and (III) the mixed-mode
state in the ($\protect\gamma $, $k_{z}$) plane. (c) The weight $c_{0}$ and
(d) skyrmion number $Q$ in the mixed mode as functions of $k_{z}$ for $%
\protect\gamma =3$ in Eq. (\protect\ref{main}) and $\Delta =0.1$, see Eq. (%
\protect\ref{Delta}). }
\label{figure4}
\end{figure}

Similar to the NN used above, parameters $o_{1-8}$ from ansatz (\ref{MM})
can be determined by means of the NN employing the Adam method to minimize
the error index, which is defined as
\begin{equation}
L\equiv\int_{\infty }^{\infty }\int_{\infty }^{\infty
}(|G_{1}|^{2}+|G_{2}|^{2})dxdy,  \label{L2}
\end{equation}%
cf. Eq. (\ref{L}), where $G_{1,2}$ are defined as%
\begin{equation}
\begin{split}
G_{1}=& \left( -\mu -\frac{1}{2}\triangledown _{\bot }^{2}\right) \psi
_{1}+(\partial _{x}-i\partial _{y})\psi _{2}+k_{z}\psi _{1} 
 -\left( |\psi _{1}|^{2}+\gamma |\psi _{2}|^{2}\right) \psi _{1}, \\
G_{2}=& \left( -\mu -\frac{1}{2}\triangledown _{\bot }^{2}\right) \psi
_{2}-(\partial _{x}+i\partial _{y})\psi _{1}-k_{z}\psi _{2} 
 -\left( \gamma |\psi _{1}|^{2}+|\psi _{2}|^{2}\right) \psi _{2},
\end{split}
\label{G1G2}
\end{equation}%
cf. Eq. (\ref{12}). The schematic of the presently used NN is shown in Fig.~%
\ref{figure4}(a). There are four layers, \textit{viz}., the input one $%
X=[1,x,y]^{T}$, the first and second hidden layers, $%
H_{l}=[1,h_{1},h_{2},h_{3},\cdots ,h_{n}]^{T}$ and $%
S_{l}=[1,s_{1},s_{2},s_{3},\cdots ,s_{n}]^{T}$, and output one $%
O=[o_{1},o_{2},o_{3},o_{4},\cdots ,o_{8}]^{T}$. The hidden and output layers
can be written as
\begin{equation}
H_{l} =[1,f(W_{1}X)^{T}]^{T},S_{l}=[1,f(W_{2}H_{l})^{T}]^{T}, 
O =Af(VS_{l}).
\label{H2}
\end{equation}%
where $W_{1}$, $W_{2}$, and $V$ are $3\times n$, $(n+1)\times n$ and $%
(n+1)\times 8$ parameter matrices, respectively.

The numerical calculation was performed for $\Delta =0.1$ [see Eq. (\ref%
{Delta})]. After $\simeq 10^{4}$ iterations, the error index (\ref{L2}) is
pushed down to $\sim 10^{-6}$. According to the distribution of weight
parameter $c_{0}$ [see Eq. (\ref{cc})], we plot the phase diagram of the GS
in the ($\gamma $,$k_{z}$) parameter plane, as shown in Fig.~\ref{figure4}%
(b), where I, II, IV and V correspond to the semi-vortex of the BV type, and
III represents the mixed mode. We have found that, similar to Ref. \cite%
{PhysRevE.89.032920}, the mixed mode exists only at $\gamma >1$, and the
phase-transition curve between the mixed mode and semi-vortex of the BV type
can be fitted by expression
\begin{equation}
\gamma =8k_{z}^{2}+4|k_{z}|+1.  \label{gamma(k)}
\end{equation}%
As an example, Fig.~\ref{figure4}(c) presents weight $%
c_{0}$ as a function of $k_{z}$ at $\gamma =3$, where the two phase
transition points are $k_{z}=\pm 0.31$ (note that the respective value $%
\left\vert k_{z}\right\vert =0.31$ is essentially smaller than $k_{z}=1$ at
which the BM $\rightarrow $ MBV phase transition occurs in the
semi-vortices, as shown above). The corresponding density distributions in
each component are shown in the first and second columns of Fig.~\ref%
{figure5}. The prediction of spatial distribution of atoms for each component 
can be tested by the Stern-Gerlach experiment.

\begin{figure}[tbp]
\centering
%Requires \usepackage{graphicx}
\includegraphics[width=4.3in]{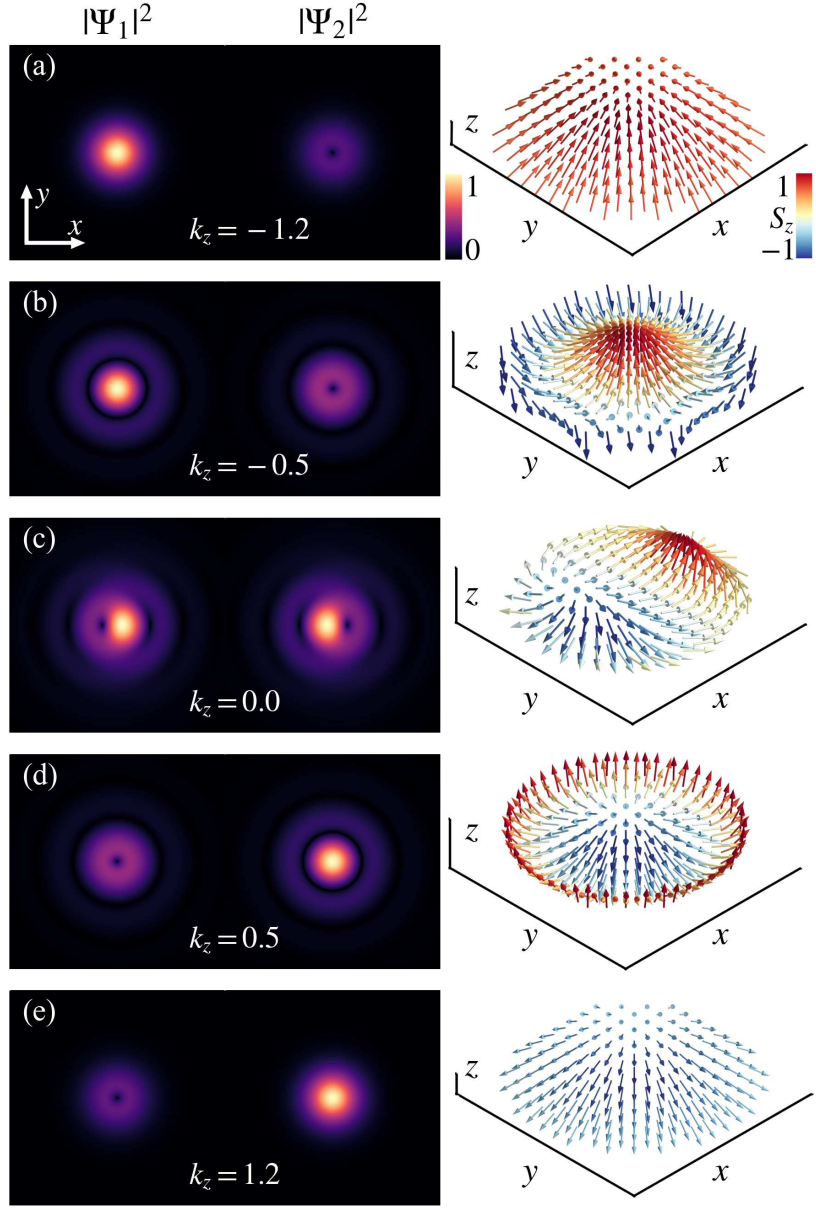}
\caption{(Color online) Density distributions of the $\protect\psi _{1}$ and
$\protect\psi _{2}$ components (the first and second columns, resectively.
The third column displays the corresponding spin textures for $\protect%
\gamma =3$ and $\Delta =0.1$. Arrows indicate the direction of the spin, and
their colors represent the component orthogonal to the ($x,y$) plane, that
is, the respective directions vary from vertical up (red) to vertical down
(blue).}
\label{figure5}
\end{figure}

For the semi-vortex of the BV type, with excitation number $m$ [see Eq. (\ref%
{vortex})], the spin vector produced by the linear solution is
\begin{equation}
\mathbf{S}_{m}=\left[\sin \Phi _{m}(r)\cos \theta ,\sin \Phi _{m}(r)\sin \theta
,\cos \Phi _{m}(r)\right],  \label{Sm}
\end{equation}
where
\begin{equation}
\sin \Phi _{m}(r)=2R_{1}R_{2}/\left( R_{1}^{2}+R_{2}^{2}\right) ,\cos \Phi
_{m}(r)=\left( R_{2}^{2}-R_{1}^{2}\right) /\left( R_{1}^{2}+R_{2}^{2}\right)
.  \label{sincos}
\end{equation}
Note properties of this vector: $\mathbf{S}_{m}=-\mathbf{S}_{-m-1}$, and $%
\lim\limits_{r\rightarrow 0}\Phi _{m}(r)=-\pi \mathrm{sign}(m)=-\Phi (r_{0})$%
, where $r_{0}$ is the minimal positive root of $R_{1}(r)R_{2}(r)=0$ [and we
set $\mathrm{sign}(0)=1$].

The SO coupling in the BEC Hamiltonian is tantamount to the
Dzyaloshinskii-Moriya interaction, supporting topologically nontrivial spin
textures, i.e., skyrmions~\cite{PhysRevLett.108.185301}. This texture is
characterized by the topological skyrmion number
\begin{equation}
Q=\frac{1}{4\pi }\iint_{\Sigma }\mathbf{S}\cdot \left( \frac{\partial
\mathbf{S}}{\partial x}\times \frac{\partial \mathbf{S}}{\partial y}\right)
dxdy,  \label{N}
\end{equation}%
where the integration domain is $\Sigma :0\leq r\leq r_{0}$, $0\leq \theta
<2\pi $, with $r_{0}$ defined above. The skyrmion number of the semi-vortex
of the BV type with excitation number $m$ is given by
\begin{equation}
Q_{\mathrm{SV}}^{(m)}=(1/4\pi )\cos \Phi (r)|_{r=0}^{r=r_{0}}\theta
|_{\theta =0}^{\theta =2\pi }=-\mathrm{sign}(m).  \label{Qm}
\end{equation}
The mixed mode can be considered as a weighted superposition of a skyrmion
and an antiskyrmion (which is sometimes called \textquotedblleft
skyrmionium\textquotedblright\ \cite{skyrmionium}), thus the respective
total skyrmion number is
\begin{equation}
Q_{\mathrm{MM}}=c_{0}Q_{SV}^{(0)}+c_{-1}Q_{SV}^{(-1)}=1-2c_{0}.  \label{Q}
\end{equation}

The spin textures of the BV-semi-vortex and mixed-mode states are summarized
in Fig.~\ref{figure5}, where the above-mentioned boundary value is $%
r_{0}=2.40/\sqrt{1-k_{z}^{2}}$ at $|k_{z}|<1$. The spin textures of the BV
semi-vortex exhibit the N\'{e}el skyrmion structure, while the MBV
semi-vortex with $k_{z}>1$ and mixed mode do not form skyrmions.

We have calculated the skyrmion number numerically, as shown in Fig.~\ref%
{figure4}(d), where the skyrmion number is correctly approximated by Eq.~%
\eqref{Q}. In particular, the fact that $Q$ takes integer values $\pm 1$ for
semi-vortices of the BV type in areas II and IV of the figure agrees with
the above-mentioned fact that only these modes form true skyrmions.

Lastly, all the GSs mentioned above are stable. This conclusion is confirmed
by direct simulations, as well as by the computation of eigenvalues for
small perturbations, cf. the stability analysis performed for 2D localized
states in other models \cite%
{PhysRevE.94.032202,Phys.Rep.303.259,Volkov,PhysicaD}.

Compared with the traditional numerical method (imaginary-time propagation method) 
for solving Gross-Pitaevskii equation, NN method has both advantages and disadvantages. 
In terms of efficiency, the NN method is time-consuming if there is no hardware acceleration.
By employ a CPU (Intel Core i7-1165G7), NN method takes 28.3 hours to obtain a 2D solution, 
while traditional numerical method only takes 7.2 hour to obtain the same solution. 
However, NN method can be accelerated by a GPU (NVIDIA RTX-2060) so that the 
calculation time can be reduced to 2.4 hours. The greatest advantage of NN method 
is that both ground state and excited state can be obtained as long as the correct 
ansatzs are given, while the traditional numerical method can only obtain the ground state. 
However, it is difficult to give the correct ansatzs, especially for a unfamiliar system. 
Note that the ansatzs~(\ref{bessel}) and~(\ref{MM})
given in this paper are based on the linear exact solution~(\ref{psi_1}).

\section{Conclusion}
We have investigated stationary solutions in the 2D model of the binary BEC,
including the SO (spin-orbit) coupling of the Rashba type and Zeeman
splitting. By introducing the generalized momentum operator, the linear
version of the system can be solved exactly. The solutions are a pair of the
BVs (Bessel vortices) or MBVs (modified Bessel vortices), in the cases of
the weak and strong Zeeman splitting, respectively. The corresponding
stationary localized vortex solutions of the full nonlinear system are
constructed by means of specially designed NN (neural network). While the
system is similar to those addressed in earlier works, a new result is the
exactly identified quantum phase transition between the semi-vortex GSs
(ground states) of the BV and MBV types. In the presence of the nonlinear
interaction between the two components of the wave function, the GS of the
mixed-mode type is constructed too, by a pair of BV states with excitation
numbers $m=0$ and $-1$, is also considered. The spin texture of the GS of
the BV-semi-vortex type exhibits the N\'{e}el skyrmion structure, and its
skyrmions number is calculated analytically, while the semi-vortices of the
MBV type and mixed modes do not form skyrmions. All the GSs identified in
this work are stable, which implies that they have the potential to be
realized in the experiment. The analysis presented here may by readily
extend to other SO-coupled systems, such as three-component BEC with the
spin-1 composition.

This research was supported by 111 project (grant No. D18001), the Hundred
Talent Program of the Shanxi Province (2018), the National Key R\&D Program 
of China under grants No. 2021YFA1400900, 2021YFA0718300, 2021YFA1400243, 
NSFC under grants Nos. 61835013 and by the Israel Science
foundation (grant No. 1286/17).

\section*{CRediT authorship contribution statement}
Huan-Bo Luo: Methodology, Writing - original draft. B. A. Malomed: Validation, Writing - review and editing. 
Wu-Ming Liu: Validation, Writing - review and editing. Lu Li: Conceptualization, Supervision, Writing - review and editing.

\end{document}